\documentclass[aps,prl,twocolumn]{revtex4}
\usepackage{epsfig}

\begin{document}

\title
{Utilizing hidden Markov processes as a new tool for
experimental physics}
\author
{Ido Kanter, Aviad Frydman and Asaf Ater}
\address{
Minerva center and the Department of Physics, Bar Ilan University,
Ramat-Gan 52900, Israel}

\begin{abstract}
A hidden Markov process is a well known concept in information
theory and is used for a vast range of applications such as speech
recognition and error correction.  We bridge between two
disciplines, experimental physics and advanced algorithms, and
propose to use a physically oriented hidden Markov process as a
new tool for analyzing experimental data.  This tool enables one
to extract valuable information on physical parameters of complex
systems. We demonstrate the usefulness of this technique on low
dimensional electronic systems which exhibit time dependent
resistance noise. This method is expected to become a standard
technique in experimental physics.
\end{abstract}

\newpage
\maketitle

In a typical scenario of experimental physics, the experiment is
designed  to detect a desired set of physical parameters. The
resultant physical data often consists of time-dependent
measurements and includes time series data. It has been long
recognized that stochastic and non-stochastic noise may provide
important insight to the physics of the system and the processes
responsible for the data \cite{noise}.

The description of complex systems suffers from the limitations of
both theoretical and experimental tools. On one hand, the
many-body system in many cases is too complicated to be solved
using existing analytical tools.  On the other hand, the
experimental results do not necessarily provide enough information
to identify the sources and the amplitudes of the possible noises,
and the physical interpretation of the outcome data stream is in
question. Hence, the emergence of a new analysis of experimental
physical data is of a great interest. We suggest a new technique
for analyzing experimental physical data. The new tool is a
physically oriented Baum-Welch (BW) algorithm \cite{2} for the
analysis of Hidden Markov Process (HMP). We bridge between two
disciplines, experimental physics and advanced algorithms, to
create a new way to study experimental findings.

A Markov model is a finite state machine that changes state once
every time unit. The manner in which the state transitions occur
is probabilistic and is governed by a state-transition matrix, M.
If, in a Markov model, the state sequence that produced the
observation sequence is not known deterministically, then the
Markov process is called a HMP. Thus, HMPs are double embedded
stochastic processes.

To exemplify a HMP in the realm of physical systems, let us assume
that a snapshot is taken from a simulation of a 1D Ising model
with nearest neighbor interactions and temperature $T$. The
snapshot consists of a sequence of $\pm 1$ elements.  The observed
snapshot is that of the pure Ising model simulation with
additional random independently flipped spins from $1$ to $-1$
(from $-1$ to $1$) with a given probability $f_+$ ($f_-$).  An
observer knows only that the physical system is 1D Ising with
nearest neighbor interactions, but he does not know the coupling
strength, $J$, nor does he know $f_{\pm}$. Hence, it is very
difficult to identify how many of the spin flips are due to a
usual thermal fluctuations and how many are a result of other
unknown processes. The observed sequence can be seen as a result
of a HMP. The elements of the HMP include the following
ingredients, where the numbers in the parenthesis stands for the
above Ising case: the number of the states in the model $q(2)$,
the Markov transition matrix M $q\times q (2\times 2)$, the
observation sequence transition distribution N $q \times q$
($f_{\pm}$), the initial state distribution of the first element
$\pi$ (a vector of rank $q(2)$) and an observed sequence of length
$L$. The aim of the BW algorithm is to accurately estimate M, N
and $\pi$ from the observation of the sequence $L$.  In our toy
model, the 1D Ising system, the physical question is to estimate
the coupling strength $J$ and the flipping rate $f_{\pm}$ from the
observation of the distorted snapshot.

An attempt to solve the above problem in its direct way is
computationally infeasible even for relatively small values of $q$
and $L$ (for instance, for $q=5$ and $L=100$, $10^{72}$ operations
are required to accurately estimate N and M\cite{2}).  A more
efficient procedure to solve this task is the BW
algorithm which is based on forward-backward procedures for the
estimation of the transition probabilities of the two stochastic
processes \cite{2}.  The key point of the BW algorithm is a
recursive calculation of the transition probabilities from the
left/right direction of each element in the sequence. For each
such element there are only finite number, $q$, possible states
independent of the length of the sequence $L$.

This approach is different than the typical scenario of the study
of physical systems which is based on the following chart flow:
{\it System's Hamiltonian} $\rightarrow$ {\it free energy} $
\rightarrow$ {\it macroscopic physical properties}. Here we
suggest the {\it reversed paradigm}. We start from the
experimental data and by utilizing the BW algorithm we try to
reveal the relevant physical properties: {\it Experimental data}
$\rightarrow$ {\it algorithm} $\rightarrow$ {\it relevant physical
properties and processes}.

\begin{figure}
\centerline{\epsfxsize=3.2in \epsffile{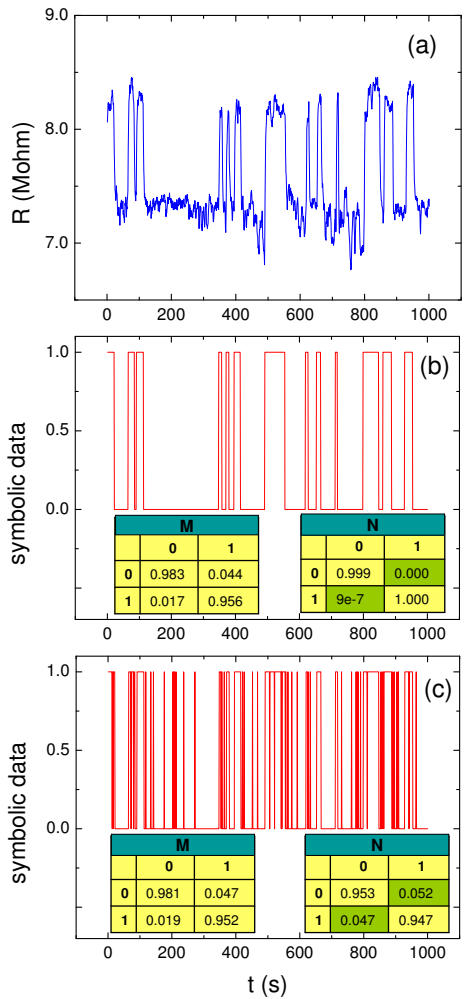}}
\vspace{1.7cm} \caption{ Resistance as a function of time, $R(t)$,
taken at T=4.2K, of a narrow $InO_x$ strongly disordered wire
prepared by step edge technique \cite{stepedge}. The wire
dimensions are $5\mu m$ long, $50nm$ wide and $30nm$ thickness.
Panel (a) shows the real data and panel (b) shows the clipped
symbolic sequence with the transition matrices of the BW analysis
for the Markov (M) and the noise (N). Panel (c) depicts the
symbolic sequence with artificially added $5\%$ noise. Note that
the relevant elements in matrix N detect this induced noise. }
\vspace{-0.4cm}
\end{figure}

As a test case for the analysis we concentrate on mesoscopic
(low-dimensional electronic) systems (for reviews see
\cite{noise,meso,meso2}). Many low dimensional systems are
characterized by large time-dependent fluctuations of the
electronic transport properties due to the fact that a relatively
small number of conductance modes govern the transport. A typical
example is shown in Fig. 1a which depicts the time dependent
resistance of a strongly disordered narrow wire. The conductance
exhibits large time dependent variations which take the form of
switching between different values. A natural interpretation of
such data is the existence of two major conductance modes between
which the conduction electrons switch due to thermal fluctuations
(also known as a "telegraph noise"). In order to check this
assumption we first generate a symbolic sequence that consists of
two levels denoted as 0/1, by clipping the sequence around 7.8
(see Fig 1b). Next we run the BW procedure on the clipped
sequence. The outcome of the BW algorithm consists of the
following two transition matrices. The first one represents the
transition probability due to the predicted Markov process between
the current state of the sequence (0 or 1) and the proceeding one.
Hence we end up with a 2x2 matrix denoted by M, where $M(i,j)$
stands for the transition probability from $j$ to $i$ . The second
matrix stands for the frequency of unexpected transitions which
are a result of a hidden process, known as the noise matrix, N.
The two matrices are shown in Fig. 1b. The interpretation of the
two matrices is as follows. The elements of the matrix M stand for
known physical properties, in our case barrier heights and the
degeneracy of the levels, while matrix N provides information
about the unknown physical processes, which can be a result of
external fields, thermal drifts, time dependent processes etc.

In Fig. 1b the off-diagonal (highlighted) elements of matrix N are
practically zero, indicating that the sequence is generated by a
simple Markov process.  The pure Markov process, {\it without
noise}, represents the fact that the main process behind these
data results from a simple Two Level System (TLS), related to the
two dominant conductance modes.  The energy levels are denoted by
$E_0=0$ and $E_1$, and the barrier from $1 \rightarrow 0$ ($0
\rightarrow 1$) is denoted by $\Delta$ ($E_1+\Delta$). The
off-diagonal elements, $M(1,0)$ and $M(0,1)$, are equal to
$\exp(-\beta(\Delta+E_1))$ and $\exp(-\beta E_1)$, respectively.
From the values of these off-diagonal elements one can derive
$\Delta$ and $E_1$. In our case $\Delta=2.6meV$ and $E_1=1.14meV$.

In order to check the reliability of the analysis we introduce
{\it artificial} random noise by flipping bits in the symbolic
sequence from $0\rightarrow 1$ or from $1 \rightarrow 0$ with
probability $f$. The data of Fig. 1b with additional $5\%$ of such
noise and the two relevant matrices are presented in Fig. 1c.  The
results of the BW algorithm show that indeed the noise is revealed
by the algorithm, while \emph{the Markov matrix is only slightly
affected}. We repeated this analysis for several data streams,
some having much larger length, and obtained similar results. It
is important to note that by eye-balling the data it is impossible
to deduce that one of these symbolic data sets is generated by a
simple Markov process while the other includes a hidden process.
Moreover, it would be hard to guess that the origin of these two
sequences is the same Markov process. This demonstrates the
potential ability of the algorithm to provide better insight into
the relevant physical processes governing the behavior of this
many-body system.

\begin{figure}
\centerline{\epsfxsize=3in \epsffile{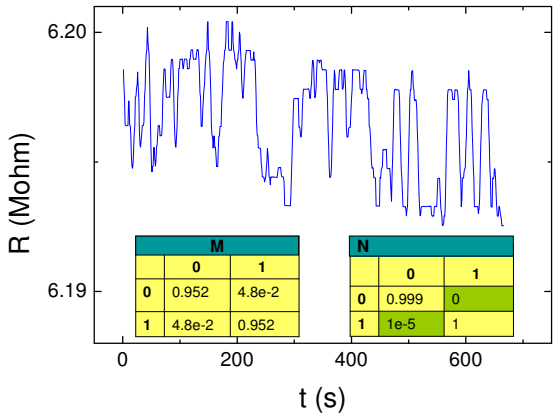}}
\vspace{0.1cm}
\caption{$R(t)$ at T=4.2K of a narrow Ni wire having dimensions of
$50\mu m$ by $50nm$ by $30nm$. The data were clipped at 6.197,
however, the resulting matrices were insensitive to the precise
clipping value} \vspace{-0.4cm}
\end{figure}

Another example of the usefulness of the analysis procedure can be
seen in Fig. 2. Here we show the data sequence of a metallic Ni wire. In
this case the data appears to be much noisier than that of Fig. 1
and a physical interpretation based on a TLS scenario is less
obvious. The experimentalist might regard this as useless physical
data. The BW analysis, on the other hand, reveals that this
sequence is also generated by a pure Markov process. The reason
why this data might appear as random noise is due to the fact
that in this case $E_1=0meV$ and $\Delta=1.1meV$.  Hence, the
barrier is less than a half of that of Fig. 1, and the two levels
are degenerate resulting in frequent transitions \cite{others}.
This example illustrates the ability of the procedure to extract
physical meaning to experimental results which might appear worthless.

In the previous examples we did not deal with physical systems
were a HMP was inherent to the system (practically zero
off-diagonal elements of N). We now provide an example for such a
scenario. Fig. 3 depicts the time dependent resistance of a dilute
2D granular Ni sample, while sweeping a magnetic field back and
forth between 2 and -2 Tesla. Application of an external field on
such samples causes sharp resistance changes at specific magnetic
fields \cite{amitay} which are superimposed on the usual time
dependent resistance sequence. In the current sample these changes
occur at fields of $-0.255$, $-0.86$, $0.97$ and $0.266~ T$.
Though the origin of these switches is not fully understood they
are very reproducible and do not originate from random telegraph
noise. The noise matrix, N, obtained from the BW procedure,
reveals considerable noise, illustrated by the ratios
$N(0,1)/M(0,1)$ and $N(1,0)/M(1,0)$ which can exceed $0.2$.  This
means that a non-negligible fraction of the transitions does not
arise from a simple Markovian process (TLS). We note that in the
absence of a magnetic field, the measured data generates Markov
matrices with practically zero noise \cite{generate}. Similar
results were obtained when allowing samples, that generate pure
Markov matrices, to drift in temperature from $4K$ to $300K$.
Clearly, the procedure is able to detect non-Markovian
perturbations to the physical systems.

\begin{figure}
\centerline{\epsfxsize=2.5in \epsffile{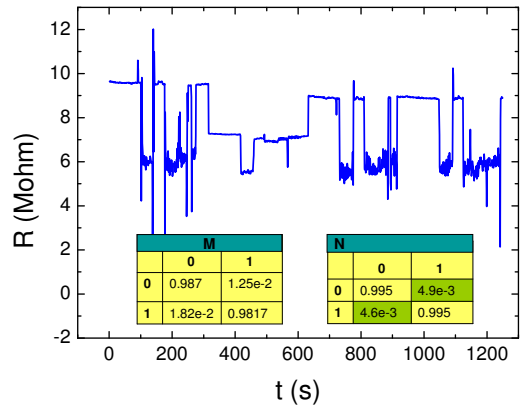}}
\vspace{0.7cm}
 \caption{ Resistance as a function of time of a
granular Ni film on the verge of electric percolation prepared by
quench condensation \cite{granularbob}.  The geometry of the
sample is a monolayer of grains with lateral dimensions of
$200nm*200nm$ and the average grain size is $10nm$. The
measurements were performed while sweeping a magnetic field back
and forth between $H=2T$ and $H=-2T$. The data was clipped at
$7.9$. } \vspace{-0.4cm}
\end{figure}

The above examples illustrate the existence of three different
regimes, characterized by the typical ratio, $R$, between the
off-diagonal elements of N and M $\{N(i,j)/M(i,j)\}$
\cite{entropy}. In the first regime, $R$ is practically zero as
exemplified in Fig. 1b.  In this case the physical generator of
the sequence is well defined.  In the second regime (as in Fig. 3)
$R$ is smaller than $1$ but finite. The conclusion should be that
the sequence is generated by a well defined Markov process,
however, some unknown processes are present and slightly influence
the data. In the third regime $R \sim 1$ (as in Fig. 1c). This
should be taken as a red alert for the experimentalist indicating
that the assumption that the data is generated from a simple
process is problematic, since most of the information is an
outcome of unknown physical processes.

In the above examples we demonstrated the effectiveness of the analysis of
HMPs in interpreting experimental data for a number of simple
mesoscopic systems. This idea can be extended to much broader
fields of application, hence we turn to a more general discussion
of the basic concepts of this bridging.

Looking more carefully at Fig. 3 one might ask himself whether it
would not be more appropriate to treat the data as being a result
of 3 levels or perhaps even more, in which case the physical
interpretation of the experimental data would change. This leads
to a more general question of how to map the experimental data to
a symbolic sequence. The main two issues are the following: (a)
What is the most appropriate vocabulary size to describe the
system, i.e. to how many levels should one clip the data? (b) For
a given vocabulary size, what are the preferred clipping
thresholds? Let us first exemplify this issue by studying the data
of Fig. 1a.  Assume that the vocabulary size is two (consisting of
0/1) and let us clip the data at 9 or 6. It is clear that the
clipped sequence consists of only zeros or only ones,
respectively. Such a sequence does not provide any insight about
the physical processes (see green boxes in Fig. 4). The obvious
reaction of an experimentalist to such an analysis is that this
unintelligent clipping suppresses all important changes in the
data. This is indeed the answer of information theory, since the
information stored in the clipped data is zero. Hence the
threshold should be chosen between the two levels. Practically, we
suggest to use one of the known clustering methods to fit the data
to two Gaussians (or, for a general case, a distribution function
for each level) representing the energy levels (centers and
broadening) \cite{duda}. The clipping is thus determined by
assigning each data point to the most probable Gaussian source.

Choosing the appropriate vocabulary size is less trivial. If prior
physical knowledge exists, it should be used to select the correct
number of levels.  If such prior knowledge is not available one
has to select the vocabulary size that would maximize the total
entropy of the HMP, $S_{HMP}(M,N)$ \cite{SHMP}. One can convince
himself that such a maximum exists by regarding the following
limiting cases. Consider again the data of Fig. 1a and let us clip
the data to an infinite number of levels (practically, $L$ levels
as the number of data points). In such a limiting scenario, the
Markov transition matrix consists of a single one in each column
and zeros elsewhere. This analysis is useless since each point
represents a switch, and the important physical switches are not
visible.  More precisely, the entropy of the generated sequence
from the Markov process is zero, and represents only the sequence
itself. Also choosing only one symbolic level results in zero
entropy. Hence, it can be expected that there is an intermediate
vocabulary size which maximizes the entropy, while providing a
stationary solution for the HMP. For instance, a similar HMP
solution should be obtained for the first/last half of the
sequence. Note that it is probable that the entropy has a number
of maxima or even a plateau rather than a unique maximal value.
For instance, in the data of Fig. 1a one can easily convince
himself that choosing three or four levels (and the appropriate
thresholds between $7.3$ and $8$) results in a similar entropy
value, since the additional levels are redundant. Practically, for
finite sequences and the experimental errors one has to define a
sensible tolerance for the entropy and a plateau is naturally
obtained.  In such a case it makes sense to choose the minimal
number of level within the plateau of maximal entropy.

The complete suggested scheme of the data analysis is thus the
following: Acquire the data sequence and assume a logical number
of levels. Determine the clipping thresholds by clustering the
data. Then, apply the BW procedure on the clipped sequence, obtain
the relevant matrices and estimate the entropy of the HMP,
$S_{HMP}(M,N)$. Repeat this procedure for different vocabulary
sizes and choose the vocabulary size that generates the maximal
Hidden Markov entropy provided that it represents a stationary
solution. The general prescription of the analysis procedure is
illustrated in Fig. 4.

\begin{figure}
\centerline{\epsfxsize=2.5in \epsffile{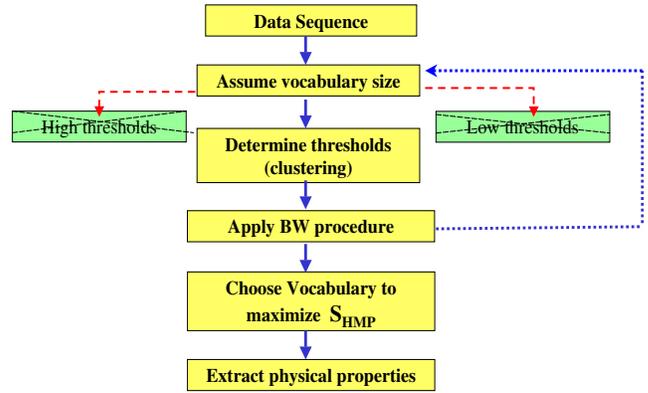}}
\vspace{1.0cm} \caption{A schematic diagram of the analysis
process.} \vspace{-0.4cm}
\end{figure}
Finally we note that the usefulness of the above-mentioned procedure
can be generalized to more complex physical systems and in particular,
systems with more than two levels and different sources for
generating switches which manifest themselves as HMPs. Such an
analysis would require a more comprehensive computational algorithm in
order to optimize the physical requirements taking into account
careful choice of thresholds and vocabulary sizes. This tool can be
applied to other physical phenomena, such as shot-noise, radioactive
decay, fluctuations in optic emission experiments etc, and is expected
to become a standard technique in experimental physics.

We are grateful to P. W. Anderson for useful discussions, to Z.
Ovadyahu for help with the measurements and H. Kfir  for technical
assistance.


\begin{references}
\bibitem{entropy} In cases where the ratio between off-diagonal
elements of matrices M and N, significantly varies, one can define
R as the ratio between the entropy of the Markov process, $-\sum_j P_j
\sum_i M(i,j)\log(M(i,j))$, and the entropy of the noise, $-\sum_j P_j
\sum_i N(i,j)\log(N(i,j))$, where $\{P_i\}$ is the stationary solution
of the Markov process.
\end{references}
\end{document}